\documentclass[%
 reprint,
superscriptaddress,
 amsmath,amssymb,
 aps, singlecolumn
]{revtex4-1}

\usepackage{graphicx}

\begin{document}

\title{Soliton dual comb in crystalline microresonators}

\author{N. G. Pavlov}
\affiliation{Moscow Institute of Physics and Technology, 141700 Dolgoprudny, Russia}
\affiliation{Russian Quantum Center, 143025 Skolkovo, Russia}
\author{G. Lihachev}
\affiliation{Russian Quantum Center, 143025 Skolkovo, Russia}
\affiliation{Faculty of Physics, M. V. Lomonosov Moscow State University, 119991 Moscow, Russia}
\author{S. Koptyaev}
\affiliation{Samsung R\&D Institute Russia, SAIT-Russia Laboratory, Moscow, 127018, Russia}
\author{E. Lucas}
\affiliation{{\'E}cole Polytechnique F{\'e}d{\'e}rale de Lausanne (EPFL), CH-1015 Lausanne, Switzerland}
\author{M. Karpov}
\affiliation{{\'E}cole Polytechnique F{\'e}d{\'e}rale de Lausanne (EPFL), CH-1015 Lausanne, Switzerland}
\author{N. M. Kondratiev}
\affiliation{Russian Quantum Center, 143025 Skolkovo, Russia}
\author{I. A. Bilenko}
\affiliation{Russian Quantum Center, 143025 Skolkovo, Russia}
\affiliation{Faculty of Physics, M. V. Lomonosov Moscow State University, 119991 Moscow, Russia}
\author{T. J. Kippenberg}
\affiliation{{\'E}cole Polytechnique F{\'e}d{\'e}rale de Lausanne (EPFL), CH-1015 Lausanne, Switzerland}
\author{M. L. Gorodetsky}
\affiliation{Russian Quantum Center, 143025 Skolkovo, Russia}
\affiliation{Faculty of Physics, M. V. Lomonosov Moscow State University, 119991 Moscow, Russia}

\begin{abstract}
We present a novel compact dual-comb source based on a monolithic optical crystalline MgF$_2$ multi-resonator stack. The coherent soliton combs generated in two microresonators of the stack with the repetition rate of 12.1 GHz and difference of 1.62 MHz provided after heterodyning a 300 MHz wide radio-frequency comb. Analogous system can be used for dual-comb spectroscopy, coherent LIDAR applications and massively parallel optical communications.
\end{abstract}

\maketitle

Kerr frequency combs \cite{DelHaye:07,Science:11} combine unique properties inherent to both narrow-linewidth lasers and broadband light sources. They enable high repetition rates in the multi-GHz to THz domain, broad octave spectrum \cite{Kippenberg:11,Gaeta:11}, and low-noise RF beat note \cite{Kippenberg:12,Weiner:11,Diddams:13}. A major advance has been the discovery of the dissipative soliton formation regime in Kerr frequency combs in nonlinear crystalline \cite{Kippenberg:14}, silica \cite{Vahala:15}, and chip-integrated Si$_3$N$_4$ optical microresonators \cite{Kippenberg:16}. This process, that has been demonstrated in a wide variety of microresonator platforms recently, enables fully coherent broadband comb operation and attract significant research interest as possible compact and high repetition rate alternatives to traditional optical frequency combs that revolutionized high-precision spectroscopy \cite{Hall&Haensch}. One promising application of frequency combs is dual comb based spectropscopy, a method that enables spectroscopy without the use of diffractive elements. The dual-comb approach \cite{Schliesser:05} allows direct conversion of optical spectra to the radio-frequency domain, and has found applications in such different areas as laser ranging \cite{Newbury:09} with sub-micron accuracy \cite{CleoKoos:14} or high-resolution spectroscopy \cite{Swann:10}.  In addition the dual comb method can be applied to coherent anti-Stokes Raman spectroscopy (CARS) \cite{Hansch:13} providing very fast Raman spectrum measurements ($\sim100$~$\mu$s) and real-time high-resolution spectral imaging. With continued development, the dual-comb spectrometer could replace traditional Fourier spectroscopy in many applications owing to its higher sensitivity \cite{Swann:10}, fast measurement and stability due to the absence of moving parts. Dual-comb spectroscopy was also demonstrated using quantum cascade lasers (QCLs) in the mid-infrared range ($\lambda \sim 4-9~\rm$~$\mu$m), appropriate for molecular rotational-vibrational absorption spectroscopy \cite{Faist:14,Faist:15}. 

Recently, dual Kerr frequency comb generation for dual-comb spectroscopy was demonstrated using a pair of optical microresonators in mid- \cite{Yu:16} and near-infrared regions \cite{Lipson:16,Vahala:16}. In Ref. \cite{Lipson:16} the  dual-comb source consisted of two silicon chip integrated microrings with slightly different radii independently coupled to the same bus waveguide. Integrated microheaters were used for accurate tuning of resonance frequencies of both resonators for combs generation using a single laser as a pump source. In dual comb spectroscopy, the free spectral range (FSR) difference $\delta \!f$ between the two generated combs defines the interferogram refresh time $\sim 1/\delta \!f$ and hence signal-to-noise ratio of the radiofrequency spectrum \cite{Hansch:13}. Well matched resonators having low difference of FSRs would provide downconversion of optical spectrum to radiofrequency range with a factor $\sim\delta\! f/f_\mathrm{FSR}$ and allow measurements using a low-frequency photodetector. It is challenging, however, to minimize the FSR difference due to the required sub-micron accuracy of microresonators manufacturing. In \cite{Vahala:16} the FSR matching of two silica resonators was possible by precise lithographic control during the manufacturing of chemically-etched wedge-resonators and using heaters. The downconversion of a $4$~THz segment of optical spectrum centered at a wavelength $\lambda \sim 1550$~nm to a RF range of $500$~MHz was demonstrated and soliton dual comb spectroscopy in a gas cell was demonstrated with a $22$~GHz resolution. 

Well matched microresonator based dual Kerr combs were also used to realize massively parallel coherent detection in telecommunication experiments \cite{EcosKoos:16}, as well as for generating interlaced channels to demonstrate 50 Tbit/s telecommunication transmission bandwidth \cite{CleoKoos:16}.

In this Letter, we present a compact optical dual-comb source based on ultra-high Q-factor crystalline MgF$_2$ optical whispering gallery mode (WGM) microresonators pumped by continuous wave lasers which may be used for compact spectroscopy or LIDAR applications. Although crystalline microresonators were the first microresonator platform in which dissipative solitons were generated, which have been applied already for creating an RF to optical link \cite{Jost:15}, achieving dual comb spectroscopy is challenging due to the requirements on the repetition rates. Here it is shown that resonators with almost identical FSR can be fabricated on the same crystalline preform. This enables the FSR of the microresonators $\sim 12.1$~GHz with a difference of only $1.62$ MHz. The demonstrated dual comb source allows the downconversion of an optical soliton frequency combs spanning $30$~nm ($3.7$~THz) around $1554$~nm to a narrow span of $300$~MHz in the RF domain.

\begin{figure}[ht]
\begin{minipage}[ht]{1\linewidth}
\center{\includegraphics[width=0.9\linewidth]{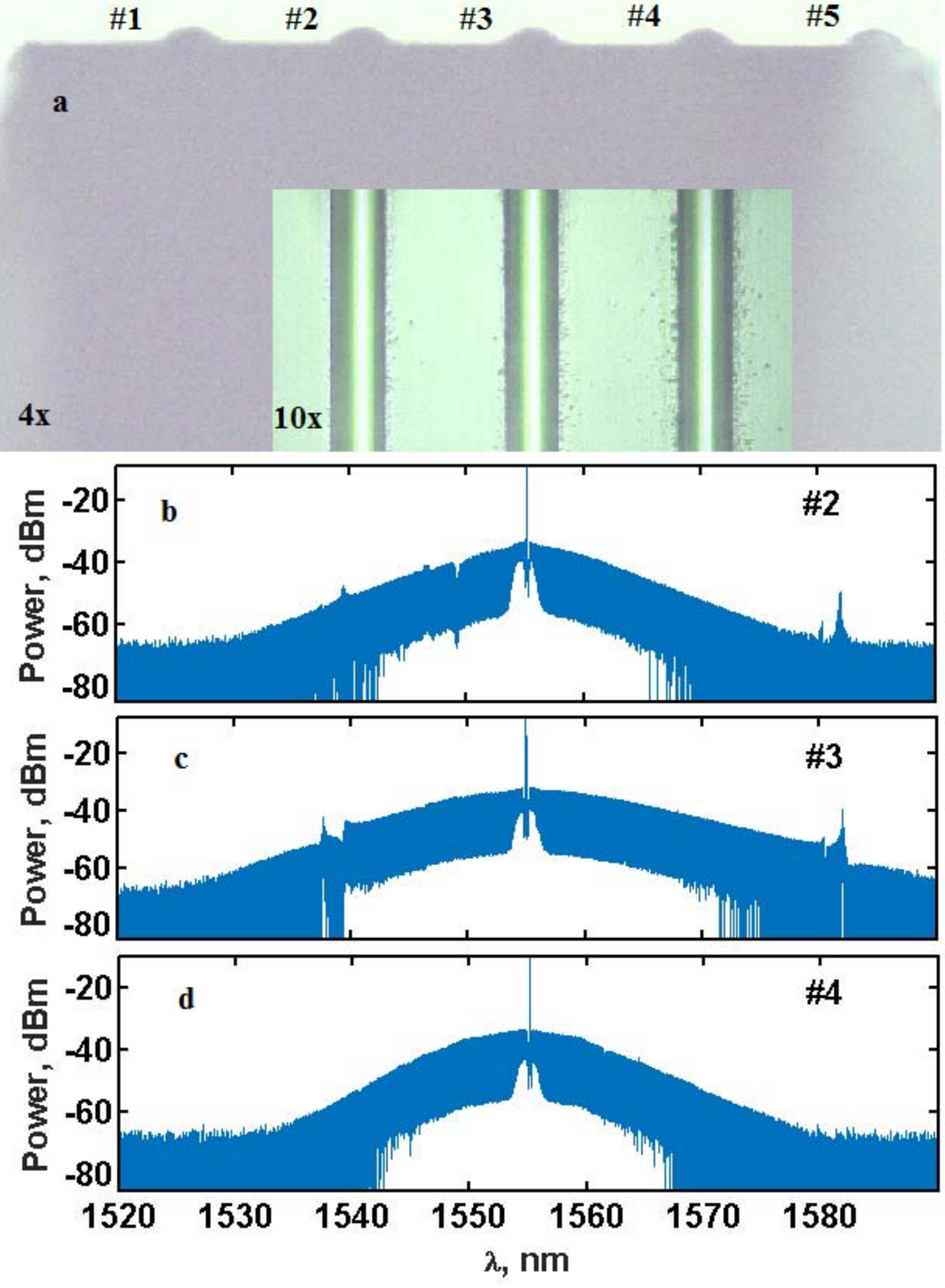}}

\end{minipage}
\caption{ (a): A view of the WGM cavities on a single crystalline rod with a major diameter of $5.68$~mm, minor diameter of $35$~$\mu$m, and distance between adjacent protrusions of $140$~$\mu$m. The inset shows $3$ cavities (\#2 -- \#4) before polishing where solitons were observed. (b-d): optical single soliton spectra generated from $3$ different cavities (\#2 -- \#4). The soliton Kerr frequency combs spans $30 - 65$~nm with a line spacing of $12.1$~GHz;}
\label{ris:image1}
\end{figure}

For a simultaneous generation of several soliton frequency combs with almost identical repetition rate, we developed a structure consisting of several identically shaped magnesium fluoride microresonators cut on the same crystalline rod as shown in Fig.\ref{ris:image1}. The structure had $5$ equal protrusions with a curvature radius of $35$~$\mu$m and a diameter of $5.68$~mm corresponding to a FSR $\sim 12.1$~GHz. The distance between adjacent protrusions was $140$~$\mu$m. This stack of microresonators was manufactured by single-point diamond turning technique \cite{Tanabe:16}, using an ultra-precise computer-controlled lens lathe (DAC ALM, DAC International, Inc.).

The quality factor after diamond cutting was approximately 10$^6$. The ultra-high intrinsic Q-factor exceeding $10^9$ was achieved by asymptotic polishing with diamond slurries \cite{Maleki:01}. To preserve the initial accuracy of the manufacturing, lint-free wipes with polishing diamond slurries were controllably applied with the same pressure to three chosen adjacent protrusions simultaneously (numbers $2 - 4$ in Fig.\ref{ris:image1}(a)). As a result of fine polishing, the difference between the FSR of several protrusions was below $10$~MHz indicating a difference of radii at the level of $0.5 - 1$~$\mu$m assuming excitation of similar mode types, sufficient for dual comb spectroscopy applications.
The loaded Q-factors of the three specially polished microresonators were ~10$^9$ after the smallest grain size polishing. The Q-factors of the other protrusions less affected by polishing procedure were $10^8$ and below.

\begin{figure}[ht]
\begin{minipage}[ht]{1\linewidth}
\center{\includegraphics[width=1\linewidth]{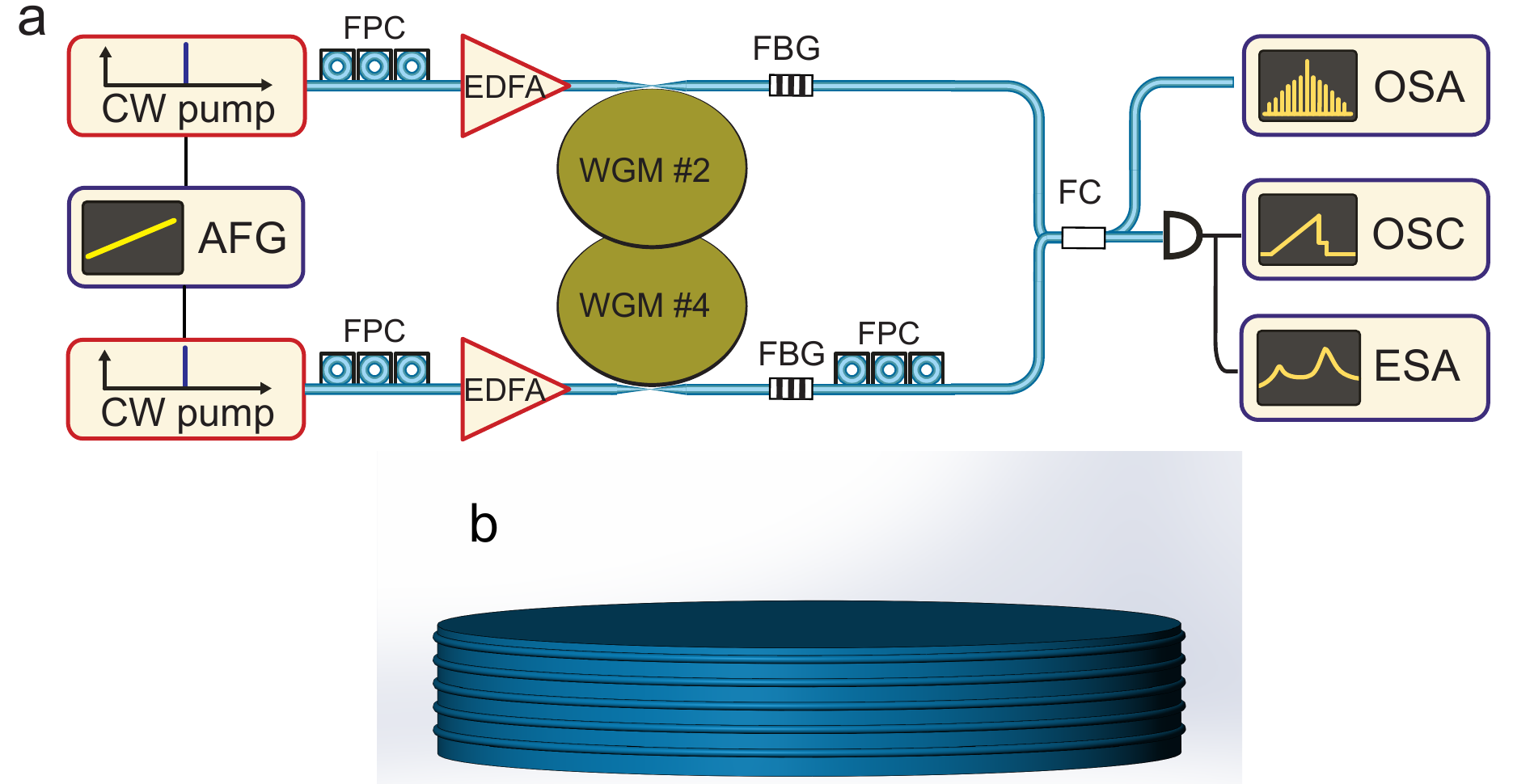}}
\end{minipage}
\caption{(a): Experimental setup to generate the soliton dual comb RF beating from the protrusions \#2 and \#4. CW: continuous-wave narrow linewidth tunable fiber laser; EDFA: erbium-doped fiber amplifier; AFG: arbitrary function generator; FPC:  fiber polarization controller; FBG: fiber Bragg grating; FC: fiber coupler; PD: photodetector; ESA: electrical spectrum analyzer; OSA: optical spectrum analyzer; OSC: oscilloscope; (b): A 3D model of the 5 microresonator stack.}
\label{ris:image2}
\end{figure}

A schematic view of the experimental setup is shown in Fig.\ref{ris:image2}. Two continuous-wave tunable narrow linewidth fiber lasers ($\lambda \sim 1554$~nm) were amplified with erbium-doped fiber amplifiers and coupled into the WGM resonators via two tapered optical fibers \cite{Hofer:10}. Each tapered fiber approached a distinct resonating protrusion (\#2 and \#4) from opposite sides and excited solitons with different FSRs. An arbitrary function generator was used to control the soliton generation in both microresonators via frequency detuning of the lasers \cite{Kippenberg:14}. Fiber polarization controllers were added to optimize the coupling efficiency. A fiber Bragg grating  was employed to suppress the transmitted pump power in the out-coupled optical signals. The repetition rates of the soliton pulses were monitored with a fast photodiode (25 GHz bandwidth) and an electrical spectrum analyzer. An optical spectrum analyzer recorded the optical comb spectra from both microresonators. The piezo-voltage of the frequency swept lasers and the transmitted optical power for both protrusions, showing the characteristic step pattern, indicating the soliton formation, were monitored with an oscilloscope.

\begin{figure}[ht]
\begin{minipage}[ht]{1\linewidth}
	\center{\includegraphics[width=0.8\linewidth]{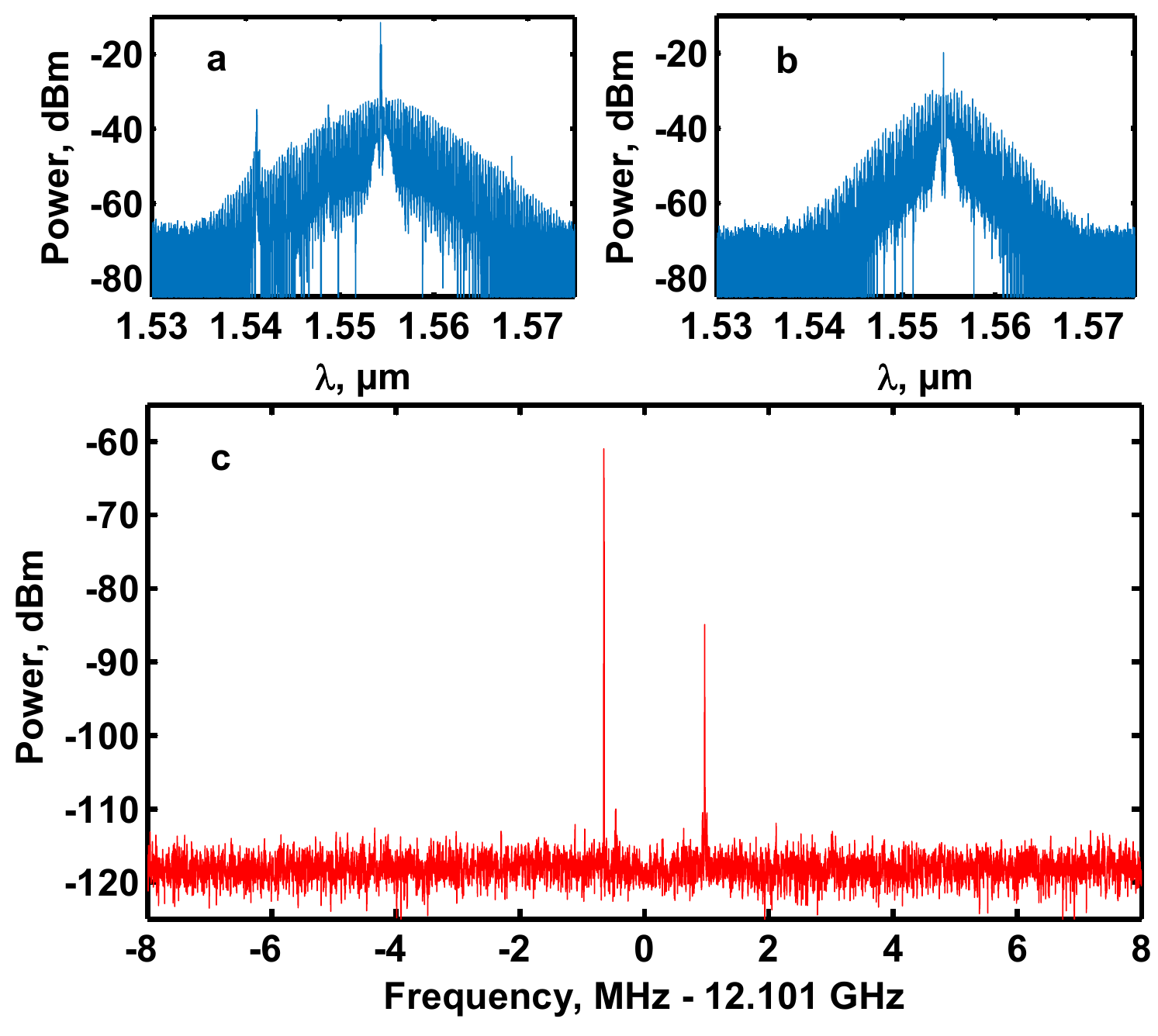}}
\end{minipage}
\caption{(a-b): Optical multisoliton spectra generated separately from cavities $\#2$ and $\#4$; (c): FSR beatnote spectrum of two simultaneous multisoliton states in two different cavities used for dual comb generation, FSR difference is $1.62$ MHz. ESA resolution bandwidth $1$ kHz.}
\label{ris:image3}
\end{figure}

In our 5-cavity structure, Kerr soliton frequency combs were observed in three different neighboring microresonators [Fig.\ref{ris:image1}]. The width of the optical soliton spectra shown in Fig.\ref{ris:image1}(b) -- Fig.\ref{ris:image1}(d) covered 30 to 65~nm around $1554$~nm. The difference between laser pump wavelengths used to excite different resonators was $8.9-30$~pm. Two soliton trains having different repetition rates $\Delta \mbox{FSR} = \mbox{FSR}_1 - \mbox{FSR}_2 = 1.62$~MHz were simultaneously generated from two different protrusions of the crystalline rod \#2 and \#4 [ Fig.\ref{ris:image1}(a)], and then combined using a fiber splitter.

To generate the dual soliton comb, we used the laser tuning method first proposed in \cite{Kippenberg:14}. We applied the ramp voltage to both lasers and tuned into the soliton state simultaneously in both cavities. The ramp starting frequency of the lasers were adjusted by aligning the characteristic soliton steps in the transmission signals from both cavities. 

\begin{figure}[ht]
\begin{minipage}[ht]{1\linewidth}
\center{\includegraphics[width=0.8\linewidth]{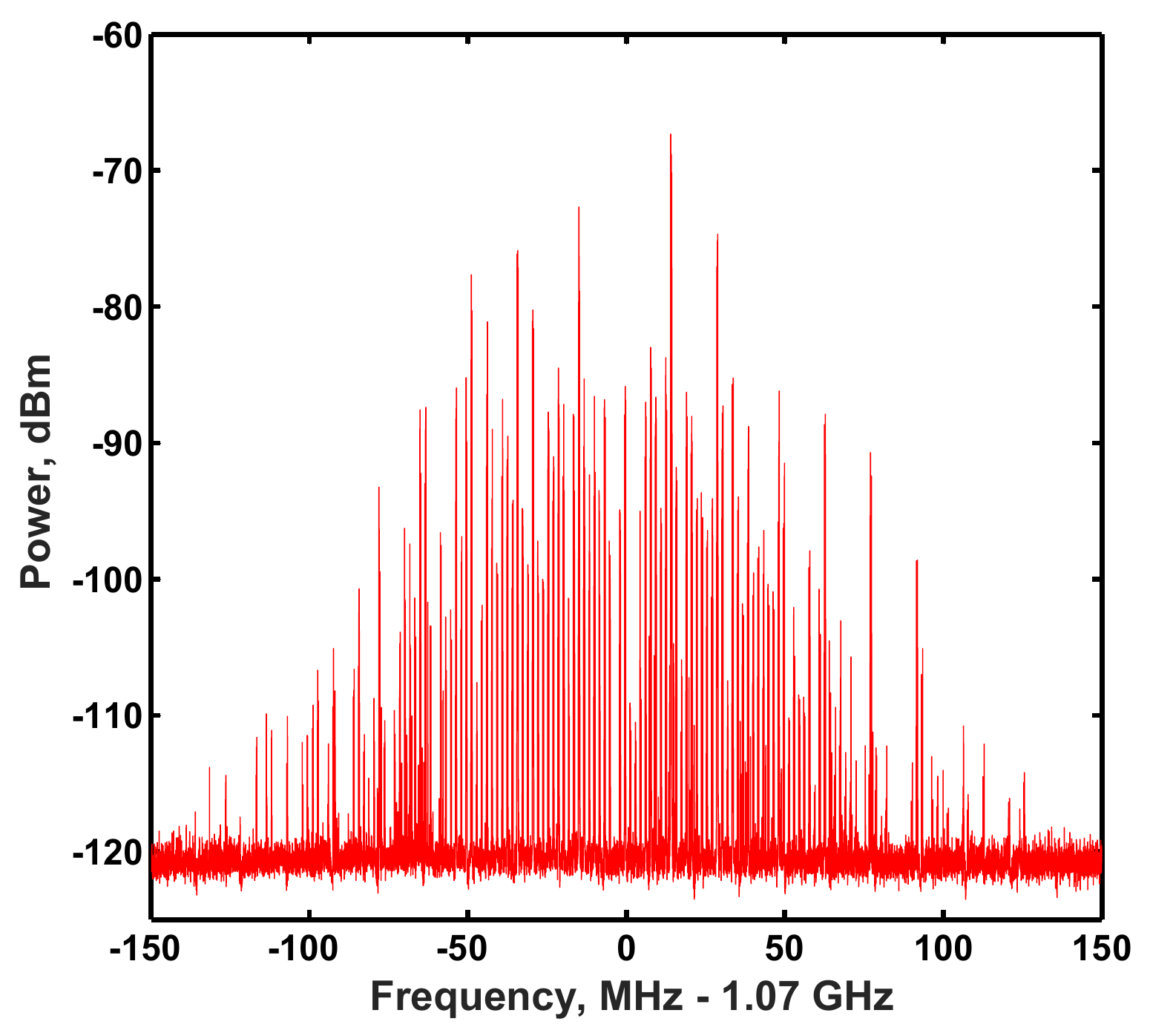}}

\end{minipage}
\caption{Radio-frequency spectrum resulting from heterodyning two multi-soliton Kerr frequency optical combs. The RF dual comb covers $300$~MHz around $1.07$~ GHz with ~$160$ lines separated by $1.62$~MHz. ESA resolution bandwidth $1$ kHz.}
\label{ris:image4}
\end{figure}

Optical spectra of multisoliton combs in both resonators are presented in Fig.~\ref{ris:image3}(a) -- Fig.~\ref{ris:image3}(b), the narrowest comb in Fig.~\ref{ris:image3}(b) contains $350$ lines separated by $12.1$~GHz and spans $35$~nm around the center wavelength $\lambda = 1554$~nm. The FSR beat note spectrum of multisoliton state combs in both resonators is shown in Fig.\ref{ris:image3}(c), FSR difference between two resonators is $1.62$~MHz and the pump lasers frequency difference is $1.07$~GHz. The dual comb downconversion from the optical domain to the RF domain [Fig.\ref{ris:image4}] results in a $300$~MHz wide RF comb, centered at $1.07$~GHz, consisting of $160$ lines spaced by $1.62$~MHz, and having a spectral envelope resulting from the optical spectrum of the two multisoliton combs. The obtained multisoliton dual comb state was short-lived ($\sim 30$ seconds), although, independently, soliton states in each cavity existed for minutes without any additional stabilization techniques. 

The short lifetime of the dual comb state is due to the thermal resonance frequency shift, which makes the presented method challenging. The soliton excitation in one cavity was observed to lead to a thermally-induced resonance frequency shift in the other cavity, thus moving its effective detuning out of the soliton existence range. As a result, we consistently obtained a soliton comb in one cavity and a high noise or MI comb state in the other cavity. 

\begin{figure}[h]
\begin{minipage}[ht]{1\linewidth}
\center{\includegraphics[width=1\linewidth]{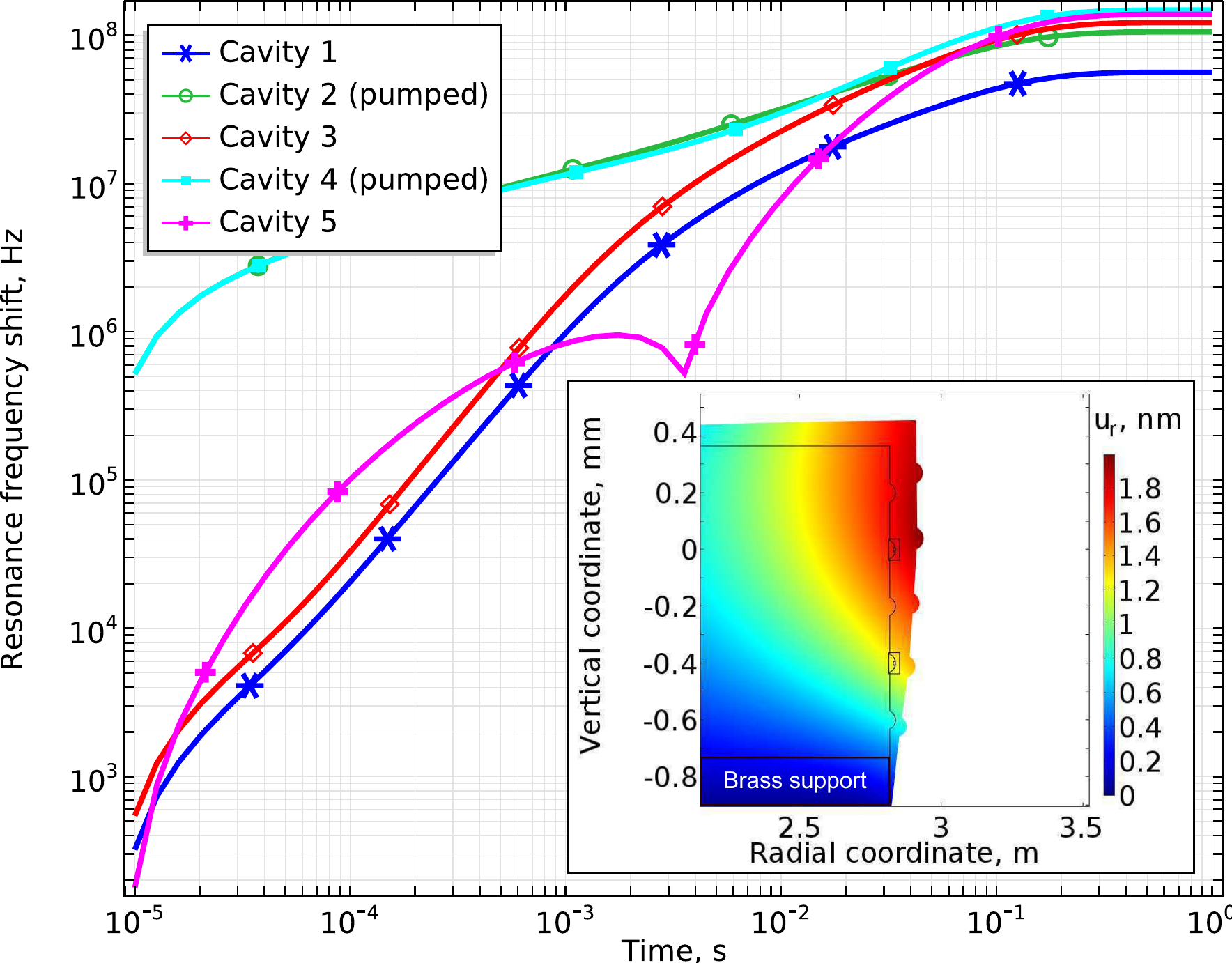}}
\end{minipage}
\caption{Numerical modeling of the resonance frequency shift induced by thermorefraction and thermal expansion of the 5-cavity system. Inset: displacement caused by heating (magnified 50000 times for visualization) in stationary regime.}
\label{ris:image5}
\end{figure}

Thermal frequency shifts in all 5 cavities were simulated using finite element method with Comsol Multiphysics software. The simulated cavity geometry coincides with the actual fabricated device: cylindrical MgF$_2$ rod is placed on a brass pedestal and cavities $\#2$ and $\#4$ are heated due to the circulating optical power. The characteristic power of 50 mW is assumed to be dissipated as heat. The material parameters were taken from the supplier site (Crystran Ltd.). We simulated the temperature variation and thermal expansion in the system assuming the heat sources were turned on simultaneously. The frequency shifts were calculated using the simple relation
\begin{align}
\frac{\Delta f}{f}=-\beta\frac{\Delta T}{n}-\frac{u_r}{R},
\end{align}
where $n$ is refraction index, $\Delta T$ is the temperature variation, $\beta$ is the thermal refraction coefficient, $u_r$ is the radial displacement of the boundary caused by thermal expansion and $R$ is the unperturbed radius. Fig.\ref{ris:image5} shows the resulting resonance frequency shifts in different cavities, the inset shows the result of the thermal expansion in stationary regime. The asymmetry in resonance shifts between cavities \#1 and \#5 at longer times is introduced by brass heat sink to which the crystal rod was glued. This heat sink was also taken into account in the simulation.  The  cavity \#5 is closer to the colder brass heat sink, and that is why less heated and deformed. The thermal expansion was found to produce over 10 times larger frequency shift than the refractive index variation. This follows from the fact that thermal expansion $\alpha$ ($u_r/R\propto \alpha \Delta T$) is much stronger in MgF$_2$ than $\beta$ (both coefficients are in fact tensors). The simulations demonstrated that pumping of one of the resonators significantly shifts the resonance frequencies of all other resonators reaching a steady state in a $1$-s timescale. In our experiment, the typical thermal frequency shifts were between $30-50$~MHz, which is in good agreement with the numerical simulation results (the resonance shift between cavities $\#2$ and $\#4$ is~$50$ MHz at $1$~s). However, the simulations  indicate that the thermal influence can be significantly reduced, by fabricating cavity protrusions separated by approximately $2$~mm or by adding a heat sink at both the top and bottom of the crystalline cylinder. The shift was also shown to depend linearly on intracavity power. In this way, we anticipate that the issues posed by the thermo-optic effects may be mitigated by appropriate design and active thermal stabilization, such as via recently developed techniques that stabilize the soliton duration \cite{Karpov:16,Yi:16} or using Pound-Drever-Hall method \cite{Kippenberg:14}. 

In conclusion: we present a precise fabrication technique of crystalline whispering gallery microresonators that allowed to achieve the FSR difference down to $\sim 1.6$~MHz. We fabricated a novel structure with several identical microresonators on a single MgF$_2$ crystalline rod with a diameter difference of only $\sim 500$~nm and demonstrated a soliton dual comb source from such structure, enabling the down conversion of $3.7$~THz of optical span centered at $\lambda \sim 1554$~nm to a $300$~MHz RF comb centered at $1.07$~GHz. Full thermomechanical modelling of the system was performed and mitigation methods for thermal effects influence were discussed. The achieved quality-factor $\sim 10^9$ and hence possible resolution is much higher than for other microresonator based platforms. Such compact dual comb sources open promising applications perspectives for spectroscopy and laser ranging.

\bigskip
{\bf Acknowledgments.}
The authors gratefully acknowledge valuable help of V.~E.~ Lobanov.

\bigskip
{\bf Funding.} Ministry of Education and Science of the Russian Federation (project RFMEFI58516X0005); Swiss National Science Foundation (grant \#163864).

\bigskip

\bibliography{Dual_comb}

\end{document}